\documentstyle[preprint,aps]{revtex}
\begin{document}
\preprint{hep-th/9410103}

\title{Action and Entropy of Extreme and Non-Extreme Black Holes}
\addtocounter{footnote}{1}

\author {Claudio Teitelboim \thanks{Electronic address:
cecsphy@lascar.puc.cl}}
\address{ Centro de Estudios Cient\'{\i}ficos de Santiago,
Casilla 16443, Santiago 9, Chile\\
and\\
Institute for Advanced Study, Olden Lane, Princeton, New
Jersey 0854, USA.}

\date{September 1994}
\maketitle

%%%%%%%%%%%%%%%%%%%%%%%%%%%%%%%%%%%%%%%%%%%%%%%%%%%%%%%%%%%%%%
\begin{abstract}
%%%%%%%%%%%%%%%%%%%%%%%%%%%%%%%%%%%%%%%%%%%%%%%%%%%%%%%%%%%%%%

The Hamiltonian actions for extreme and non-extreme black holes
are compared and contrasted and a simple derivation of the lack
of entropy of extreme black holes is given. In the non-extreme
case the wave function of the black hole depends on horizon
degrees of freedom which give rise to the entropy. Those
additional degrees of freedom are absent in the extreme case.

%%%%%%%%%%%%%%%%%%%%%%%%%%%%%%%%%%%%%%%%%%%%%%%%%%%%%%%%%%%%%%
\end{abstract}
\vspace{1cm}

It has been recently proposed \cite{hawking}, \cite{horowitz}
that extreme black holes have zero entropy \cite{wilczek}. The
purpose of this note is to adhere to this claim by providing an
economical derivation of it. The derivation also helps to set
the result in perspective and to relate it to key issues in the
quantum theory of gravitation, such as the Wheeler-De Witt
equation.

The argument is the application to the case of an extreme black
hole of an approach to black hole entropy based on the
dimensional continuation of the Gauss-Bonnet theorem developed
in \cite{btz}. The approach in question had been previously
applied to non extreme black holes only \cite{ct}.

To put into evidence as clearly as possible the distinction
between extreme and non extreme holes, we first perform the
analysis for the non-extreme case and then see how it is
modified in the extreme case.

We will deal with gravitation theory in a spacetime of dimension
$D$ with positive definite signature (Euclidean formulation). To
present the argument in what we believe is its most transparent
form for the purpose at hand, we will start with the Hamiltonian
action and will only at the end discuss the connection with the
Hilbert action.

For non-extreme black holes the Euclidean spacetimes admitted in
the action principle have the topology I$\!$R$^2 \times
S^{D-2}$.  It is useful to introduce a polar system of
coordinates in the I$\!$R$^2$ factor of I$\!$R$^2 \times
S^{D-2}$. The reason is that the black hole will have a Killing
vector field --the Killing time-- whose orbits are circles
centered at the horizon.  We will take the polar angle in
I$\!$R$^2$ as the time variable in a Hamiltonian analysis. An
initial surface of time $t_1$ and a final surface of time $t_2$
will meet at the origin. There is nothing wrong with the two
surfaces intersecting. The Hamiltonian can handle that.

The canonical action

\begin{equation}
I_{can} = \int(\pi^{ij}\dot{g}_{ij} - N{\cal H} - N^i {\cal H}_i),
\label{1}
\end{equation}
{\em without any surface terms added} can be taken as the
action for the wedge between $t_1$ and $t_2$ provided the
following quantities are held fixed:\\
(i) the intrinsic geometries $^{(D-1)} {\cal G} _1$, $^{(D-1)}
{\cal G} _2$ of the slices $t=t_1$ , and $t=t_2$,\\
(ii) the intrinsic geometry $^{(D-2)} {\cal G}$ of the $S^{D-2}$
at the origin\\
(iii) the mass at infinity, with an appropiate asymptotic
fall-off for the field.

The term ``mass'' here refers to the conserved quantity
associated with the time Killing vector at infinity. It is thus
more general than the $P^0$ of the Poincar\'e group, which only
exists when the spacetime is a symptotically flat.  For example
when there is a negative cosmological constant this mass is the
value of a generator of the anti-de Sitter group.

Note that we have listed the intrinsic geometry of the $S^{D-2}$
as a variable independent from the three-geometries of the
Slices $t=t_1$ and $t=t_2$.  This is because in the variation of
the action (\ref{1}) there is a separate term in the form of an
integral over $S^{D-2}$, which contains the variation of $^{(D-2)}
{\cal G}$.

It should be observed that there will be no solution of the
equations of motion satisfying the given boundary conditions if,
for example, one fixes the mass at $t_2$ to be different from
the mass at $t_1$.  However in the quantum theory one {\em can}
take $M_1 \neq M_2$, the path integral will then yield a factor
$\delta(M_2 -M_1)$ in the amplitude.  Similarly there will be no
solution of the equations of motion unless the geometry of the
$S^{D-2}$ at the origin as approached from the slice $t =t_1$,
coincides with the one corresponding to $t=t_2$, and unless that
common value also coincides with the one taken for the geometry
of the $S^{D-2}$ at the origin.  However these precautions
need not be taken in the path integral, which will automatically
enforce them by yielding appropiate $\delta$-functionals.  This
situation is the same as that arising with the action of a free
particle in the momentum representation, where there is no
clasical solution unless the initial and the final momenta are
equal, but yet, one can (and must) compute the amplitude to go
from any initial momentum to any final momentum.

To the action (\ref{1}) one may add any functional of the quantities
held fixed and obtain another action appropiate for the same boundary
conditions.  In particular one may replace (\ref{1}) by

\begin{equation}
I= I_{can} + B[^{(D-2)} {\cal G}],
\label{2}
\end{equation}
where $B[^{(D-2)} {\cal G}]$ is any functional of the
$(D-2)$-geometry at the origin.  If we only look at the wedge
$t_1 \leq t \leq t_2$ there is no privileged choice for $B$.
However if we demand that the action we adopt should also be
appropiate for the complete spacetime, then B is uniquely fixed.
This is because when one deals with the complete spacetime the
slices $t=t_1$ and $t=t_2$ are identified and neither $^{(D-1)}
{\cal G}_1$ nor $^{(D-1)} {\cal G}_2$ nor $^{(D-2)} {\cal G}$
are held fixed.  Now, unlike its Minkowskian signature
continuation, the Euclidean black hole obeys Einstein's
equations {\em everywhere}. Thus it should be an
extremum of the action with only the asymptotic data (mass) held
fixed.  The demand that the action should be such as to have the
black hole as an extremum with respect to variations of
$^{(D-2)} {\cal G}$ fixes

\begin{equation}
B = 2\pi A(r_+)\;\;\;\; \mbox{(non-extreme case)}.
\label{3}
\end{equation}
where $A(r_+)$ is the area of the $S^{D-2}$ at the origin.

Note that if one includes $B$ for the full spacetime one must
include it for the wedge as well. This is because (i) the full
spacetime is a particular case of the wedge, and (ii) the
boundary term (\ref{3}) depends only on the $(D-2)$ geometry at
the origin and not on $t_1$ or $t_2$,

The way in which (\ref{3}) arises is the following.  First one
writes the metric near the origin in ``Schwarzchild
coordinates'' as

\begin{equation}
ds^2 = N^2(r, x^p) dt^2 +N^{-2}(r, x^p) dr^2 +\gamma_{m n}(r,
x^p) dx^m dx^n,
\label{4}
\end{equation}
with

\begin{equation}
(t_2 -t_1)N^2 = 2\Theta(x^p)(r-r_+) + O(r-r_+)^2\;\;\;
\mbox{(non-extreme case)}.
\label{5}
\end{equation}

Here $r$ and $t$ are coordinates in I$\!$R$^2$ and $x^p$ are
coordinates in $S^{D-2}$. The parameter $\Theta$ is the total
proper angle (proper length divided by proper radius) of an arc
of very small radius and coordinate angular opening $t_2 -t_1$
in the I$\!$R$^2$ at $x^p$.  For this reason it is called the
opening angle.  When the sides of the wedge are identified $2\pi
-\Theta$ becomes the deficit angle of a conical singularity in
I$\!$R$^2$.

Next, one evaluates the variation of the canonical action
(\ref{1}) to obtain

\begin{eqnarray}
\delta I_{can} & = & -\int_{S^{(D-2)}(r_+)} \Theta (x^p) \delta
\gamma^{1/2}(x^p) d^{D-2}x + \beta \delta M  +
\nonumber \\
       &   & \int \pi^{ij} \delta g_{ij} \left|^2 _1 \right. +
           \mbox{(terms vanishing on shell)}.
\label{6}
\end{eqnarray}
Here $\beta$ is the Killing time separation at infinity.

Last, one observes that when the slices $t=t_1$ and $t=t_2$ are
identified, the term $\int \pi^{ij} \delta g_{ij} \left|^2 _1
\right.$ cancels out.  Thus if $M$ and $J$ are kept fixed
but $\gamma^{1/2}(x^p)$ is allowed to vary one must add
(\ref{3}) to (\ref{1}) in order to obtain from the action
principle that at the extremum.

\begin{equation}
\Theta (x^p) = 2\pi\;\;\;\;\;\; \mbox{(complete spacetime,
non-extreme case)}.
\label{7}
\end{equation}

Equation (\ref{7}) must hold because otherwise there would be a
conical singularity at $r_+$ and Einstein's equations would be
violated in the form of a $\delta$ -function source at the origin.

Let us now turn to the extreme case.  By definition of an
extreme black hole the square lapse $N^2$  has a double root at the
origin.  Thus one must replace (\ref{5}) by

\begin{equation}
(t_2 -t_1)N^2 =  O(r-r_+)^2\;\;\; \mbox{(extreme case)}.
\label{8}
\end{equation}

This means that one must have

\begin{equation}
\Theta (x^p) =0  \;\;\;\; \mbox{(extreme case)},
\label{9}
\end{equation}
instead of (\ref{7}).  It then follows that

\begin{equation}
B =0  \;\;\;\; \mbox{(extreme case)},
\label{10}
\end{equation}
so that the canonical action (\ref{1}) is appropiate
\underline{as is} for extreme black holes.

Note that equation (\ref{8}) holds not only for the complete
spacetime but also for a wedge of the extreme black hole
geometry.  This implies that (\ref{9}) must hold also off-shell
(for all configurations allowed in the action principle).  This
is so because for the wedge there is no way to obtain $\Theta
=0$ by extremizing the action since $^{(D-2)} {\cal G}$ is held
fixed.

The difference between non-extreme and extreme cases has a
topological origin.  For all $\Theta$'s in the interval

\begin{equation}
0< \Theta \leq 2\pi,
\label{11}
\end{equation}
the topology of the $t, r$ piece of the complete spacetime is
that of a disk with the boundary at infinity. When $\Theta <
2\pi$ the disk has a conical singularity in the curvature at the
origin with deficit angle $2\pi - \Theta$.  When $\Theta = 2\pi$
the singularity is absent.

However when $\Theta =0$ the topology is different. Indeed, what
would appear naively to be a source at the origin in the form of
a ``fully closed cone'' --as was misunderstood in \cite{btz}--
is really the signal of a spacetime with different topology. As
the cone closes, its apex recedes to give rise to the infinite
throat of an extreme black hole.  Thus the origin is effectively
removed from the manifold whose $t, r$ piece is no longer a
disk, but rather, an annulus whose inner boundary is at infinite
distance.

Now, one wants to include in the action principle fields of a
given topology so that one can continuously vary from one to
another.  Therefore for the complete spacetime of the
non-extreme case all fields obeying (\ref{11}) are allowed so
that (\ref{7}) only holds on-shell.  On the other hand, for the
extreme case we must have (\ref{9}) to also hold off-shell. This
is so since if the origin is removed, and there is no place to put a
conical singularity.

We reach therefore an important conclusion: if we demand that
the action should have an extremum on the black hole solution,
then we must use a different action for extreme and non-extreme
black holes.  This means that these two kinds of black holes are
to be regarded as drastically different physical objects, much
in the same way as particle of however small but finite mass is
drastically different from one of zero mass \cite{gibbons}.  The
discontinuous jump in the action is just the way that the
geometrical theory at hand has to remind us that extreme and
non-extreme black holes fall into different topological classes.

The action may be rewritten as

\begin{equation}
I = 2\pi \chi A(r_+) + I_{can},
\label{12}
\end{equation}
and equations (\ref{5}) and (\ref{8}) may be summarized as

\begin{equation}
(t_2 -t_1)N^2 = 2\chi \Theta(x^p)(r-r_+) + O(r-r_+)^2,
\label{12.1}
\end{equation}
where $\chi$ is the Euler characteristic of the $t, r$ factor
of the complete black hole spacetime.  For the non-extreme case
one has $\chi =1$ (disk), and for the extreme case $\chi =0$
(annulus).  Expression (\ref{12}) had been anticipated in
\cite{btz}, where it emerged naturally from a study of the
dimensional continuation of the Gauss-Bonnet theorem, but it
was missed there that $\chi =0$ corresponds to extreme black
holes.

If one evaluates the action on the black hole solution one finds

\begin{equation}
I_{can}(Black Hole) =0,
\label{13}
\end{equation}
because the black hole is stationary ($\dot{g}_{ij} =0$) and
because the constraint equations ${\cal H} = {\cal H}_i =0$ hold.
Thus one has

\begin{equation}
I(Black Hole) =2\pi \chi A(r_+).
\label{14}
\end{equation}

Now, the action (\ref{12}) is appropiate for keeping $M$ fixed.
In statistical thermodynamics this corresponds to the
microcanonical ensemble.  Thus, for the entropy $S$ in the
classical approximation one finds

\begin{equation}
S= (8\pi G \hbar)^{-1}2\pi \chi A(r_+),
\label{15}
\end{equation}
where we have restored the universal constants. Thus one sees
that extreme black holes ($\chi =0$) have zero entropy.

A word is now in place about the relation of (\ref{12}) with the
Hilbert action

\begin{equation}
I_H = \frac{1}{2} \int_M \sqrt{g} R d^Dx - \int_{\partial M}
\sqrt{g} K d^{D-1}x,
\label{16}
\end{equation}

As was shown in \cite{btz}, for the complete spacetime
(\ref{12}) and (\ref{16}) just differ by a boundary term at
infinity, which automatically regulates the divergent functional
(\ref{16}).  This assertion is not valid for the wedge.  In that
case, as was also noted in \cite{btz}, (\ref{12}) and (\ref{16})
differ not only by a boundary term at infinity but also by a
boundary term at the origin. For the complete spacetime one has

\begin{equation}
I = I_H -B_{\infty} ,
\label{16.1}
\end{equation}
whereas for the wedge

\begin{equation}
I = I_H + \pi(2\chi -1) A(r_+) -B_{\infty} - \pi A_{\infty}.
\label{16.2}
\end{equation}
For the reasons given above we adopt (\ref{12}) and not
(\ref{16}) as the action for the wedge.

The discontinuous change in the action between extreme and
non-extreme black holes has dramatic consequences for the wave
functional of the gravitational field in the presence of a black
hole --which one may call for short the wave function of the
black hole. Indeed, in the extreme case, the wave function has
the usual arguments, namely, it may be taken to depend on the
geometry of the spatial section and on the asymptotic time
separation $\beta$,

\begin{equation}
\Psi = \Psi[^{(D-1)} {\cal G}, \beta].
\label{17}
\end{equation}

The dependence of $\Psi$ on the three geometry is governed by
the Wheeler -- De Witt equation

\begin{equation}
{\cal H} \Psi = 0,
\label{18}
\end{equation}
whereas the dependence on the asymptotic time $\beta$ is governed by the
Schr\"odinger equation

\begin{equation}
\frac{\partial \Psi}{\partial \beta} + M\Psi= 0,
\label{19}
\end{equation}
where $M$ is the mass as defined by Arnowitt, Deser and Misner
(see for example \cite{ct2}). On the other hand, for the
non-extreme case the wave function has an extra argument which
may be taken to be the opening angle $\Theta$,

\begin{equation}
\Psi = \Psi[^{(D-1)} {\cal G}, \beta, \Theta].
\label{21}
\end{equation}

Since according to (\ref{6}) $\Theta$ is canonically conjugate
to $\gamma^{1/2}$, one has in addition to equations (\ref{18})
and (\ref{19}) the extra Schr\"odinger equation at the
horizon \cite{carlip}

\begin{equation}
\frac{\delta \Psi}{\delta \Theta(x)} - \gamma^{1/2}(x) \Psi =0.
\label{22}
\end{equation}

The additional horizon degree of freedom canonical pair
$(\gamma^{1/2}, \Theta)$ may be regarded as responsible for the
black-hole entropy in the non-extreme case.  Indeed there is no
entropy in the extreme case precisely because then the origin is
absent and there is no place for $(\gamma^{1/2}, \Theta)$ to sit
at.  This agrees with a point of view previously expressed
\cite{carlip}, namely that, --in a way yet to be spelled-- the
black-hole entropy could be conjectured as arising from
``counting conformal factors on the $S^{D-2}$ at $r_+$'' or, in
terms of the canonically conjugate statement ``from counting
two-dimensional geometries within a small disk at the horizon''.
That disk is removed in the extreme case --and with it the entropy.

ACKNOWLEDGMENTS

The author is very grateful to J. Zanelli for many
inlightening discussions and for much help in preparing this
manuscript.  Thanks are also expressed to Dr. A. Flisfisch for
his kind interest in the author's work. This work was supported
in part by Grant 194.0203/94 from FONDECYT (Chile), by a
European Communities contract, and by institutional support to
the Centro de Estudios Cient\'{\i}ficos de Santiago provided by
SAREC (Sweden), and a group of chilean private companies (COPEC,
CMPC, ENERSIS, CGEI). This research was also sponsored by IBM
and XEROX-Chile.

\end{document}